# Level-1 Regional Calorimeter Trigger System for CMS

P. Chumney, S. Dasu, J. Lackey, M. Jaworski, P. Robl, W. H. Smith
*University of Wisconsin, Madison WI, 53706, USA*

The Compact Muon Solenoid (CMS) calorimeter regional trigger system is designed to detect signatures of isolated and non-isolated electrons/photons, jets, τ-leptons, and missing and total transverse energy using a deadtimeless pipelined architecture. This system contains 18 crates of custom-built electronics. The pre-production prototype backplane, boards, links and Application-Specific Integrated Circuits (ASICs) have been built and their performance is characterized.

## 1. INTRODUCTION

The Compact Muon Solenoid (CMS) is a general-purpose detector that will operate at the Large Hadron Collider (LHC). Its construction is currently underway at the European Laboratory for Particle Physics (CERN) near Geneva, Switzerland. This large detector will be sensitive to a wide range of new physics at the high proton-proton center of mass energy $\sqrt{s}$ =14 TeV [1].

At the LHC design luminosity of $10^{34}$ cm$^{-2}$ s$^{-1}$, a beam crossing every 25 ns contains on average 17.3 events. These $10^9$ interactions per second must be reduced by a factor of $10^7$ to 100 Hz, the maximum rate that can be archived by the on-line computer farm. This will be done in two steps. The level-1 trigger first reduces the rate to 75 kHz, and then a High Level Trigger (HLT), using an on-line computer farm, handles the remaining rate reduction.

The Compact Muon Solenoid level-1 electron/photon, τ-lepton, jet, and missing transverse energy trigger decisions are based on input from the level-1 Regional Calorimeter Trigger (RCT) [2]. The RCT plays an integral role in the reduction of the proton-proton interaction rate ($10^9$ Hz) to the High Level Trigger input rate ($10^5$ Hz) while separating physics signals from background with high efficiency. The RCT receives input from the brass and scintillator CMS hadron calorimeter (HCAL) and PbWO$_4$ crystal electromagnetic calorimeter (ECAL), that extend to |η|=3 (pseudorapidity η is –ln(tan(θ/2) where the x-y plane is perpendicular to the beam line defined by z). An additional hadron calorimeter in the very forward region (HF) extends coverage to |η|=5. A calorimeter trigger tower is defined as 5x5 crystals in the ECAL of dimensions 0.087x0.087 (ΔφxΔη), which corresponds 1:1 to the physical tower size of the HCAL. Since the HF is not used in any electron or photon algorithm, it has a coarser segmentation in η and φ.

The algorithm to find electron and photon candidates uses a 3x3 calorimeter trigger tower sliding window centered on all ECAL/HCAL trigger towers out to |η|=2.5. Two types of electromagnetic objects are defined, a non-isolated and an isolated electron/photon. Four of each type of electron per regional crate are forwarded to the Global Calorimeter Trigger (GCT) for further sorting. The top four candidates of each type are received by the level-1 Global Trigger (GT).

The jet trigger uses the transverse energy sums ($E_{T,ECAL}+E_{T,HCAL}$) for each 4x4 trigger tower calorimeter region in the barrel and endcap [3]. In the very forward region (3<|η|<5) of the CMS spectrometer, each HF tower is treated as a single region and their Δφ division matches that of the 4x4 regions of the barrel and endcap. The jet or τ-tagged jet is defined by a 3x3 region $E_T$ sum. In the case of τ-tagged jets (only |η|<2.5), none of the nine regions are allowed to have more than 2 active ECAL or HCAL towers (i.e. above a programmable threshold). Jets in the HF are defined as forward jets. Four of the highest energy central, forward, and τ-tagged jets are selected, allowing independent sorting of these 3 jet types until the final stage of jet sorting and trigger decision. In total, there are twelve jets available at the GT level.

## 2. CALORIMETER TRIGGER HARDWARE

The regional calorimeter trigger electronics comprises 18 crates for the barrel, endcap, and forward calorimeters and one cluster crate to handle the jet algorithms. These will be housed in the CMS underground counting room adjacent to and shielded from the underground experimental area.

Twenty-four bits comprising two 8-bit calorimeter energies, two energy characterization bits, a LHC bunch crossing bit, and 5 bits of error detection code will be sent from the ECAL, HCAL, and HF calorimeter electronics to the nearby RCT racks on 1.2 Gbaud copper links. This is done using one of the four 24-bit channels of the Vitesse 7216-1 serial transceiver chip on calorimeter output and RCT input, for 8 channels of calorimeter data per chip. The RCT V7216-1 chips are mounted on mezzanine cards, located on each of 7 Receiver Cards and one Jet/Summary Card for all 18 RCT crates. The eight mezzanine cards on the Receiver Cards are for the HCAL and ECAL data and the one mezzanine card located on the Jet/Summary Card is for receiving the HF data. The V7216-1 converts serial data to 120 MHz TTL parallel data, which is then deskewed, linearized, and summed before transmission on a 160 MHz ECL custom backplane to 7 Electron Isolation Cards and one Jet/Summary Card. The Jet/Summary Card receives the HF data and sends the regional $E_T$ sums to the cluster crate and the electron candidates to the GCT. The cluster crate implements the jet algorithms and forwards the 12 jets to the Global Calorimeter

Trigger. Construction of the cluster crate is pending, awaiting studies of including its function in the GCT.

The Receiver Card, in addition to receiving calorimeter data on copper cables using the V7216-1, shares data on cables between RCT crates. Synchronization of all data is done with the local clock and the Phase ASIC (Application-Specific Integrated Circuit--described below). The Phase ASIC also checks for data transmission errors. Lookup tables are used to translate the incoming $E_T$ values onto several scales and set bits for electron identification. Adder blocks begin the energy summation tree, reducing the data sent to the 160 MHz backplane.

The Electron Isolation Card receives data for 32 central towers and 28 neighboring towers via the backplane. The electron isolation algorithm is implemented in the Electron Isolation ASIC described below. Four electron candidates are transmitted via the backplane to the Jet/Summary (J/S) Card. The electrons are sorted in a Sort ASIC on the J/S Card and the top 4 of each type are transmitted to the GCT for further processing. The J/S Card also receives $E_T$ sums via the backplane, and forwards them and two types of muon identification bits (minimum ionizing and quiet bits – described later) to the GCT. A block diagram of this dataflow is shown in Fig. 1.

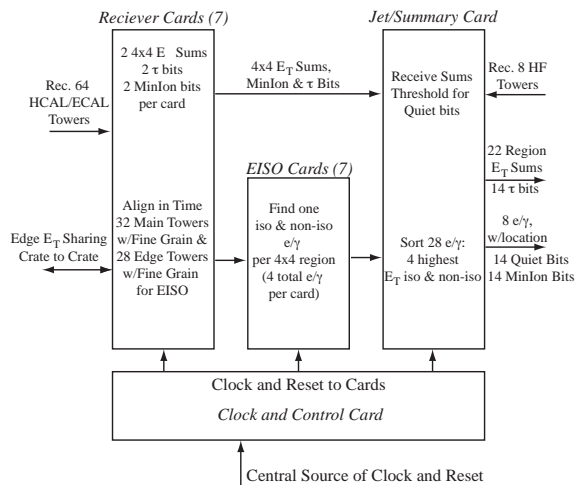

Fig. 1. Dataflow diagram for the crate, showing data received and transferred between cards on the 160 MHz differential ECL backplane. Brief explanations of the card functionality are shown. For more details see the text or reference [2].

To implement the algorithms described above, five high-speed custom Vitesse ASICs are used: a Phase ASIC, an Adder ASIC, a Boundary Scan ASIC, a Sort ASIC, and an Electron Isolation ASIC [4]. They were produced in Vitesse FX™ and GLX™ gate arrays utilizing their sub-micron high integration Gallium Arsenide MESFET technology. Except for the 120 MHz TTL input of the Phase ASIC, all ASIC I/O is 160 MHz ECL.

The Phase ASICs on the Receiver Card align and synchronize the data received on four channels of parallel data from the Vitesse 7216. The Adder ASICs sum up eight 11-bit energies (including the sign) in 25 ns, while providing bits for overflows. The Boundary Scan ASIC handles board level boundary scan functions and drivers for data sharing. Four 7-bit electromagnetic energies, a veto bit, and nearest-neighbor energies are handled every 6.25 ns by the Electron Isolation ASICs, which are located on the Electron Isolation Card. Sort ASICs are located on the Electron Isolation Card, where they are used as receivers, and are located on the J/S Cards for sorting the e/γ. In the cluster crate the Adder and Sort ASICs are used for the jet algorithms. The Adder, Phase, Boundary Scan, and Sort ASICs have been successfully tested and procured in the full quantities needed for the system. The remaining Electron Isolation ASIC has been manufactured and is installed on the prototype boards described below [4].

## 3. PRE-PRODUCTION PROTOTYPES

The successful conclusion of the first generation prototype program proved the design as described in Chapter 5 of the CMS Trigger Technical Design Report (TDR), which was approved in March 2001 [2], marked the start of the construction of full-function pre-production modules based on the TDR. A custom pre-production prototype 9U VME crate, Clock and Control Card, Receiver Card, Electron Isolation Card, and Jet/Summary Card have been produced with the above ASICs. Mezzanine cards with the Vitesse 7216-1 serial link for the Receiver Card and dedicated test cards have also been constructed.

We have built a pre-production VME crate, shown in Fig. 2, with a new backplane that implements all of the level-1 trigger algorithms approved by CMS and the LHCC, and documented in the Trigger TDR [2]. The backplane is located in the middle of the crate between a card cage 400 mm deep in the rear and a card cage 280 mm deep in the front. As shown in Fig. 2, the backplane is a monolithic, custom, 9U high printed circuit board with front and back card connectors. The top 3U of the backplane utilizes 4 row (128 pin) DIN connectors, capable of full 32-bit VME. The two leftmost front slots of the backplane use three row (96 pin) DIN connectors in the P1 and P2 positions with the standard VME pin assignments. Thus, a standard VME system module can be inserted in the two front stations with a form factor conversion between the first slot and remaining slots performed on the custom backplane.

The rear of the crate behind the two standard VME stations is occupied by the 48V power supply. In the data processing section of the crate (the bottom 6U of the backplane) a single high speed, controlled impedance, AMP 340-pin stripline connector is used for both front and rear card insertion. The stripline connectors have a power plane between every four signal pins, producing an impedance of 50Ω. Separate high current contacts, provided for the power plane connections, are used to transmit power to the boards. The connectors are housed in a cast aluminum shell that doubles as a board stiffener. The electrical characteristics of the connectors are excellent, allowing sub-nanosecond rise times with very low crosstalk. Data is transmitted across the backplane in 160 MHz differential ECL.

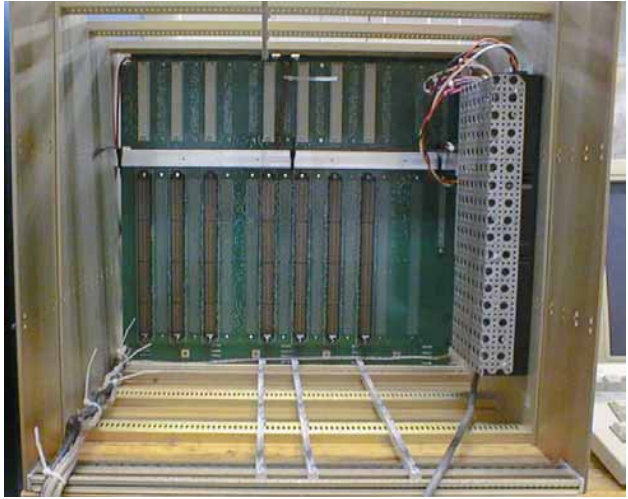

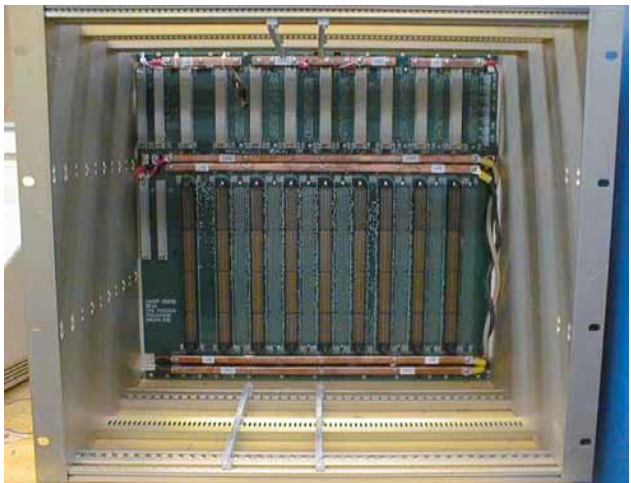

Fig. 2. Pre-production crate: rear (top picture) and front (bottom picture) view.

We have built a new Clock and Control Card, (front side shown in Fig. 3), that matches the timing in the new backplane and cards. The Clock and Control Card distributes 160 MHz clocks and resets across the backplane to all cards, with delay adjusts on the back of the card to allow for differences in travel times across the backplane. The 120 MHz clocks for the Vitesse 7216-1 serial links and the Phase ASICS are also sent out to the Receiver Cards and Jet/Summary Card via the backplane. Power distribution on all the cards is handled with DC to DC converters fed with 48 V from the backplane power pins.

We have constructed a full function Receiver Card, shown in Fig. 4 and Fig. 5, on which we have installed the Adder, Phase and Boundary Scan ASICs, as well as eight of the new version of the 4 x 1.2 Gbit/s serial receiver mezzanine card, shown in Fig. 6. The new Receiver Card features mezzanine cards with an improved version of the Vitesse serial receiver chip that is more tolerant of clock jitter. Two channels of tower data for each of four channels of serial link are received per mezzanine card for a total of 64 channels of data from the HCAL and ECAL per Receiver Card. Parallel data out of the Vitesse serial link are aligned via the Phase ASIC and then sent to the memory lookup (LUT) to linearize the data and set the electron identification bit for the electron algorithms. From the memory lookup, seven bits of ECAL $E_T$ and one electron identification bit are sent to the Electron Isolation Card via the Boundary Scan ASICs. The Boundary Scan ASICs also handle the shared data coming in from other crates via the SCSI type cables at the front of the card. In addition, a sum of $E_{T,HCAL} + E_{T,ECAL}$, is sent from the memory lookup to two stages of Adder ASICs to form the 4x4 region sums that then travel via the backplane to the Jet/Summary Card.

We have also built a dedicated new Serial Test Card, shown in Fig. 7, for testing the receiver mezzanine cards and performing detailed bit error checking. Additional copies of this card will be used for production testing of the mezzanine cards and integration tests with ECAL and HCAL electronics.

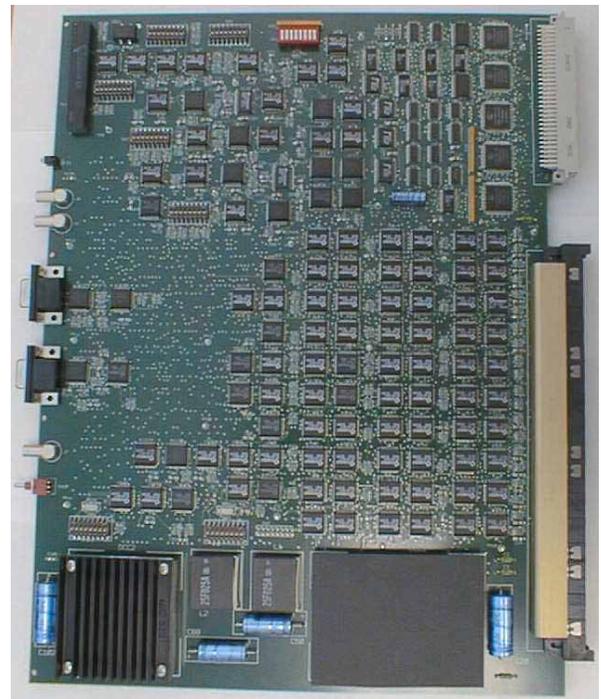

Fig. 3. Pre-production Clock & Control Card.

A full function Electron Isolation Card has been built (Fig. 8) on which we have installed the Electron Isolation and Sort ASICs. Data are received over the backplane from the Receiver Card via the Sort ASICs' differential input. The Sort ASIC's sorting feature can be set on or off and is off on the Electron Isolation Card. This data includes 16 towers for each Region and 28 towers of neighbor data coming from adjoining Receiver Cards or neighboring crates via the data sharing cables at the front of the Receiver Card. Alignment in time is done at the Boundary Scan ASIC on the Receiver Card. These 44 towers are sent to two Electron Isolation ASICs which choose the two highest energy electrons of each type, isolated and non-isolated. A memory lookup compresses the seven bit energies to six bits and sets a location bit for each region served by the card. The $E_T$s and position information for the four electrons are forwarded via the backplane to the Jet/Summary Card.

The pre-production prototype Jet/Summary Card has been manufactured and the front side is shown in Figure 9. Via the backplane it receives 14 Region sums from 7 Receiver Cards using two Sort ASICs. These are all forwarded via SCSI cable at 80 MHz to the Cluster Crate that will implement the jet-finding algorithm. Twenty-eight electron candidates are received from the Electron Isolation Card via the backplane and sorted with two Sort ASICs. Only the top four of each type, isolated and non-isolated, are forwarded to the Global Calorimeter Trigger, where all the electron candidates from the 18 RCT crates are sorted again and the top four of each type sent on to the Global Trigger. In addition, a Quiet Bit for each 4x4 Region is set depending on a programmable threshold for each 4x4 region. Fourteen Minimum Ionizing Bits, formed by an "OR" of 16 HCAL quality bits per region on the RC, are forwarded with the Quiet bits to the GCT on the Electron Cables.

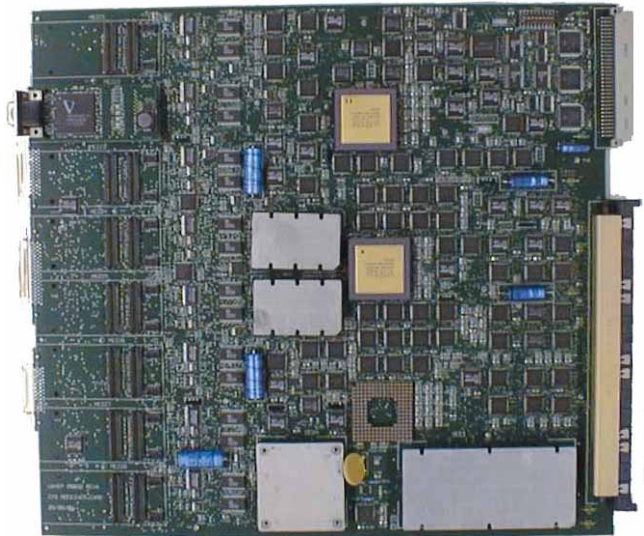

Fig. 4. Pre-production Receiver Card (front) showing Adder ASICs and a Mezzanine Receiver Card installed.

The HF data is received via one receiver mezzanine card located at the top of the J/S card (seen in Figure 9). Since each HF tower also represents one HF Region, the data are handled by the Phase ASIC, and then two Memory lookups handle the energy assignment of the four different η slices of the HF. This information is delayed using Boundary Scan ASICs to align it with the Region $E_T$ sums from the Receiver Cards and then forwarded with a quality bit to the GCT.

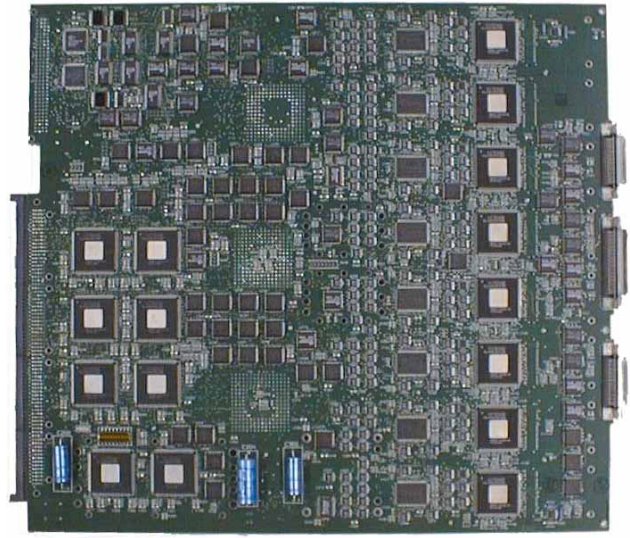

Fig. 5. Pre-production Receiver Card (rear) showing Phase and Boundary Scan ASICs and lookup tables.

## 4. TEST RESULTS

We are engaged in a detailed testing and validation program of these pre-production trigger boards. We are testing a fully functional crate with the pre-production Backplane, Receiver Card, Clock Card, Electron Isolation Card, and Jet/Summary Card. When complete, these tests will fully verify all 5 Vitesse custom ASICs. Thus far, we have validated the Phase, Sort, Boundary Scan, and Adder ASICs and completed their production. The receiver mezzanine card with the Vitesse 7216-1 transceiver chip has been validated as well. The Receiver Card, Clock Card and Electron Isolation Card have been validated and their production versions are being manufactured. The Jet/Summary Card, Crate, and Backplane will be verified using a full crate of cards. Verification of data pathways and logic function is checked using Boundary Scan, which is fully implemented on all boards and ASICs. The Electron Isolation ASIC is mostly validated. Full validation requires data from neighboring Receiver Cards, so we are currently manufacturing several more Receiver Cards and Electron Isolation Cards in order to fully test the Electron Isolation ASIC and verify the clocking and operation of a full crate. A full crate is also required for planned system integration tests with the Calorimeter in spring 2004 [5].

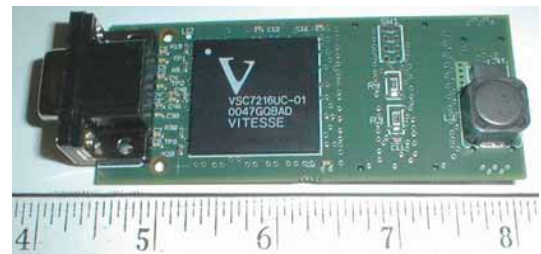

Fig. 6. Close up of a pre-production 4 x 1.2 Gbit/s Copper receiver mezzanine card.

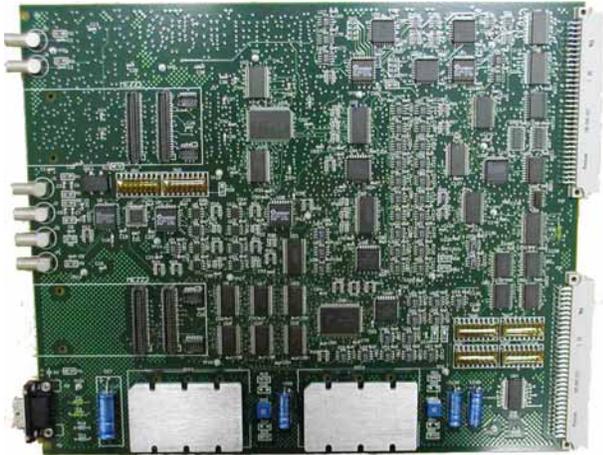

Fig. 7. Pre-production Serial Link Test Card.

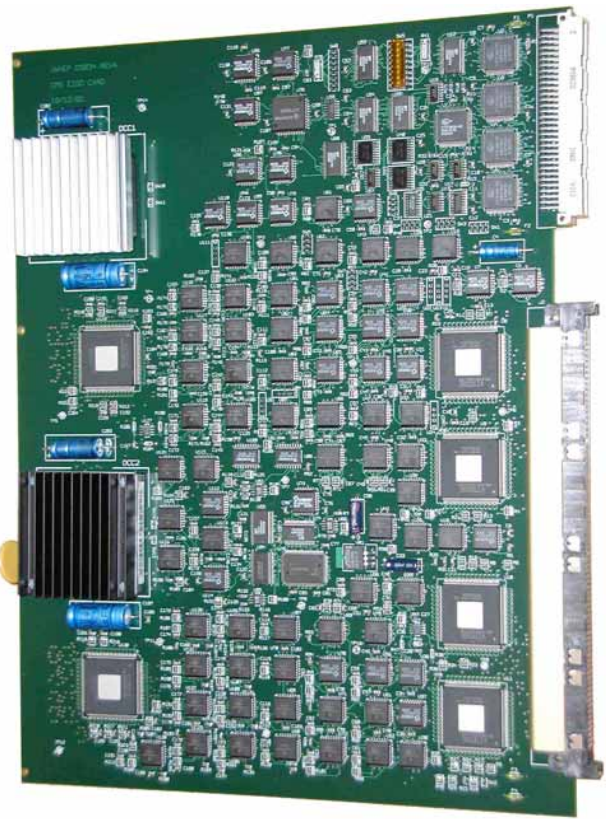

Fig. 8. Pre-production Electron Isolation Card with Electron Isolation and Sort ASICs installed.

Additionally, we have conducted tests of the pre-production 4 x 1.2 Gbit/s copper serial link system. The cable used in the tests is composed of two 20 m lengths of 22 AWG Spectra Strip Skew-Clear® shielded 2-pair cable with VGA-type 15-pin DIN connectors. We have built a special "test" transmitter mezzanine card to drive the signals over the cables and used two Serial Link Test Cards to continuously transmit and receive pseudo-random data over many days with a trap on error, yielding a bit error rate of less than $10^{-15}$. A full production run is complete and the boards will be used to test the RCT electronics in-house and sent out to other laboratories to assure compatibility of the RCT with the ECAL and HCAL electronics [5].

## 5. SUMMARY

Construction of the CMS level-1 Regional Calorimeter Trigger Pre-production Prototype Receiver Card, Isolation Card, Jet/Summary Card, Clock and Control Card, Serial Link Mezzanine Card, Backplane, Crate and five custom ASICs that implement the CMS level-1 calorimeter trigger algorithms is complete. Tests conducted thus far have validated the design of the production boards.

### Acknowledgments

This work was supported by the United States Department of Energy and the University of Wisconsin.

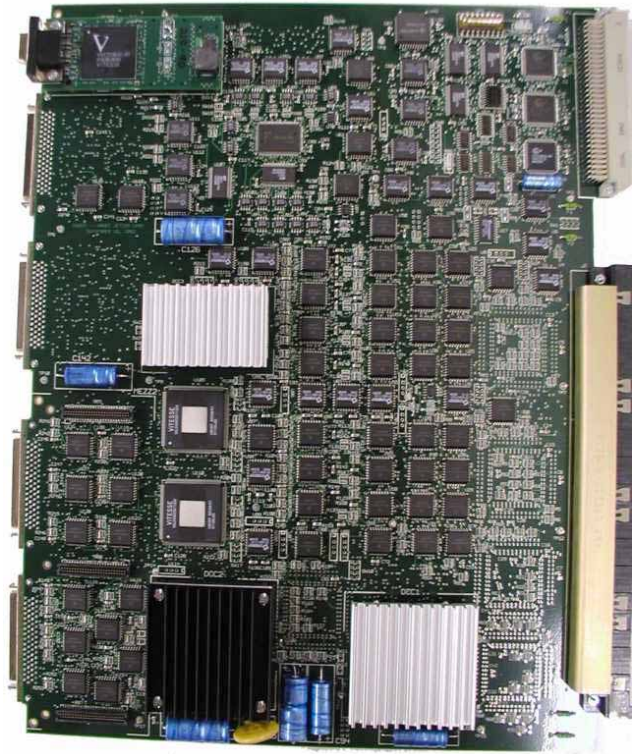

Fig. 9. Pre-production Jet/Summary Card.